\documentclass[3p,twocolumn]{elsarticle}

\usepackage{lineno,hyperref}

\usepackage{graphicx}
\usepackage{epstopdf}
\usepackage{dcolumn}
\usepackage{amsmath}
\usepackage{amsfonts}
\usepackage{amsthm}
\usepackage{amssymb}
\usepackage{mathrsfs}
\usepackage{bbold}
\usepackage{bm}

\begin{document}

\begin{frontmatter}

\title{Scaling of the local quantum uncertainty at quantum phase transitions}

%---------------------------------------
%% Group authors per affiliation:
\author{I. B. Coulamy} 
%\ead{ivanbc@if.uff.br}

\author{J. H. Warnes}
%\ead{jherazow@if.uff.br}

\author{M. S. Sarandy$^{* ,}$ \fnref{fn1}}
%\ead{msarandy@if.uff.br}
\cortext[cor1]{Corresponding author}
\fntext[fn1]{Tel.: +55-21-2629-5802 / Fax: +55-21-2629-5887 }

\author{A. Saguia}
%\ead{amen@if.uff.br}

\address{ Instituto de F\'{\i}sica, Universidade Federal Fluminense - Avenida General Milton Tavares de Souza s/n, Gragoat\'a, 24210-346,Niter\'oi, RJ, Brazil}

%------------------------------------
\begin{abstract}
We investigate the local quantum uncertainty (LQU) between a block of $L$ qubits and one single qubit in a composite system of $n$ qubits 
driven through a quantum phase transition (QPT). A first-order QPT is analytically considered through a Hamiltonian implementation of the  
quantum search. In the case of second-order QPTs, we consider the transverse-field Ising chain via a numerical analysis through density 
matrix renormalization group. For both cases, we compute the LQU for finite-sizes as a function of $L$ and of the coupling parameter, 
analyzing its pronounced behavior at the QPT. 
\end{abstract}

%--------------------------------------

\end{frontmatter}

%\linenumbers

%%%%%%%%%%%%%
\section{Introduction}
%%%%%%%%%%%%%
The interplay between quantum information theory and statistical mechanics has brought emerging connections between these 
research fields~\cite{Amico:08,Maruyama:09,Georgescu:14}. In particular, it has provided a deeper understanding about the role 
played by correlations in quantum phase transitions (QPTs). A seminal result in this direction is a link between the scaling 
of pairwise entanglement and QPTs in quantum spin chains~\cite{Amico:02,Osborne:02}. This has been further developed by 
introducing a distinction between the characterization of first-order and continuous QPTs~\cite{Wu:04,Wu:06}. For a block analysis, 
entanglement entropy has been found to be related to the central charge of the Virasoro algebra 
associated with the conformal field theory behind the critical model~\cite{Vidal:03,Korepin:04,Calabrese:04}. 
More generally, it has been shown that quantum correlation measures such as provided by the quantum discord~\cite{Ollivier:01} 
are also able to identify quantum criticality~\cite{Dillenschneider:08,Sarandy:09}. Remarkably, pairwise quantum discord may exhibit a more 
robust characterization of QPTs than pairwise entanglement in certain cases. For instance, pairwise quantum discord between distant sites in a quantum 
chain may indicate a quantum critical point, while entanglement is absent already for very short distances~\cite{Maziero:12,Huang:13}. In addition, 
for finite temperatures, pairwise quantum discord is able to reveal the QPT by non-analyticities in its derivatives, while the pronounced 
behavior in two-qubit entanglement disappears for even small temperatures~\cite{Werlang:10}. 

In this work, we aim at investigating the behavior of the local quantum uncertainty (LQU)~\cite{Girolami:13} at quantum criticality. 
The LQU has been introduced as a quantum discord-like measure, which is primarily related with the skew information~\cite{Luo:03,Wigner:63}. 
In particular, it plays a role in the characterization of quantum metrology protocols~\cite{Girolami:13,Yu:14}. The behavior of LQU between 
pairs of spins in a quantum spin chain has been recently considered~\cite{Karpat:14,Carrijo:15}. 
%%%v2
Here, we generalize this previous analysis for systems of dimension $2\times 2$ by considering 
the LQU for blocks of arbitrary dimension $D \times 2$ and also by discussing its finite-size behavior in both first-order and second-order QPTs. 
%%%
More specifically, we will evaluate the LQU between a block of $L$ quantum bits (qubits) and one single qubit in a composite system of $n$ qubits. 
For a first-order QPT, we will consider a Hamiltonian implementation of the  
quantum search, which is designed to find out a marked element in an unstructured search space of $N=2^n$ elements. By analytical evaluation, 
we will show that the LQU exponentially saturates to a constant value 
at the critical point as we increase the block length $L$. This saturation is found to be enhanced by the system size $n$. 
On the other hand, at non-critical points, the LQU will be shown to vanish for large $n$. 
In the case of second-order QPTs, we consider the transverse-field Ising 
model with open boundary conditions. By implementing a numerical analysis via density matrix renormalization group (DMRG), 
we will show that the concavity of the LQU as a function of the block size $L$ characterizes the QPT. For both first-order and 
second-order QPTs, we also consider the LQU as a function of the coupling parameter, showing that the LQU exhibits a pronounced behavior 
at the quantum critical point independently of the block sizes of $L$ qubits. In particular, this pronounced behavior is sensitive to $n$, showing a 
scaling behavior as we increase the size of the system.

%%%%%%%%%%%%%
\section{Local quantum uncertainty}
%%%%%%%%%%%%%

The uncertainty of an observable $K$ in a quantum state $\rho$ is usually quantified by the variance $V(\rho,K)=Tr\rho K^2-(Tr\rho K)^2$.
It may exhibit contributions from both quantum and classical sources.  Quantum uncertainty comes from the noncommutativity between 
$K$ and $\rho$, being quantified by the skew (not commuting) information~\cite{Luo:03,Wigner:63} 
\begin{equation}
I(\rho,K)= Tr\rho K^2-Tr\rho^{1/2}K\rho^{1/2}K.
\end{equation}
Indeed, suppose $\rho$ and $K$ commute. Then, $\rho$ and $K$ have a common basis of eigenstates $\{|k\rangle\}$, which means that 
the uncertainty of $K$ in an individual eigenstate $|k\rangle$ vanishes.  
Hence, a nonvanishing uncertainty $V(\rho,K)$ is only possible if 
$\rho$ is a classical mixing of $\{|k\rangle\}$. Therefore, the commutation 
of $\rho$ and $K$ implies that $V(\rho,K)$ has a classical origin. 

The quantum uncertainty is intrinsically connected with the concept of quantum correlation. For example, let us consider
a Bell state of two qubits, namely, $|\psi\rangle = (|00\rangle + |11\rangle)/\sqrt{2}$, where $\{|0\rangle, |1\rangle\}$ denotes the 
computational basis. This is an eigenstate of the
global observable $\sigma_{z}\otimes \sigma_{z}$, so there is no uncertainty on the result of a measurement of such an observable. 
On the other hand, the measurement of local spin observables is intrinsically uncertain for the density operator 
$|\psi\rangle \langle \psi|$, since an entangled state cannot be an eigenstate of a local observable. In particular, the 
variance $V(\rho,K)$ for a local observable $K$ will vanish if and only if the state is uncorrelated.

\begin{figure}[!ht]
\centering
\includegraphics[scale=0.3]{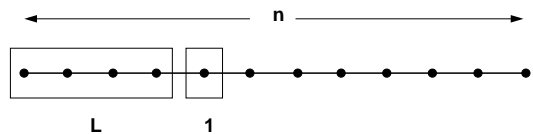}
\caption{Bipartition used to defined the subsystems $A$ and $B$ for the LQU evaluation. The size $L$ of the block $A$ is arbitrarily chosen and subsystem $B$ is taken as one qubit.}
\label{fig:chain}
\end{figure}

The concept of quantum uncertainty can be extended to mixed states. In this case, the skew information $I(\rho,K)$ vanishes if and only if $\rho$ is 
not disturbed by the measurement of $K$. If $K$ is a local observable, the states left invariant by local measurement are the states with
zero quantum discord with respect to that local subsystem~\cite{Li:08}. The quantum uncertainty on local observables is then intimately related to the notion of quantum
discord and, as shown in Ref.~\cite{Girolami:13}, it can be used as a discord-like quantifier. 
We are now ready to define the local quantum uncertainty (LQU). Let $\rho=\rho_{AB}$ be the state of a bipartite system, and let $K^{\Gamma}$ 
denote a local observable on $B$ ($K$ is represented by a Hermitian operator on $B$ with nondegenerate spectrum $\Gamma$). 
The LQU as defined in [4], is given by
\begin{equation}
 Q(\rho) = min_{K^{\Gamma}} I(\rho,K^{\Gamma}). 
\end{equation} 
Notice that $Q$ is the minimum quantum uncertainty associated to a single 
measurement on subsystem $B$.  If there is a $K$ for which $Q=0$ then there is no quantum correlation between the two parts 
of the state $\rho$. As proved in Ref.~\cite{Girolami:13}, the LQU satisfies all the good properties of a discord-like measure.
An analytical expression for $Q$ can be obtained if we consider a
bipartite $D \times 2$ system. In this case
\begin{equation}
Q(\rho_{AB}) = 1-\lambda_{max}(W_{AB}) ,
\label{LQU}
\end{equation}
where $\lambda_{max}$ is the  maximum eigenvalue of the $3\times 3$ symmetric matrix $W$ whose elements are given by
\begin{equation}
(W_{AB})_{ij}=Tr[\rho^{1/2}_{AB}( I_{A}\otimes  \sigma_{iB})\rho^{1/2}_{AB}( I_{A}\otimes  \sigma_{jB})] .
\label{eq.lqu.def}
\end{equation}
In this work, we will consider a set of $n$ qubits aligned in a chain, with the bipartition in subsystems $A$ and $B$ chosen as shown in Fig.~\ref{fig:chain}.  
\vspace{0.2cm}
\vspace{0.2cm}

%%%%%%%%%%%%%%%
\section{LQU for the quantum search}
%%%%%%%%%%%%%%%
The aim of the search problem is to find out a marked element in an unstructured list of $N$ candidates. In a quantum setting, it is possible to solve the search problem  
with scaling $\sqrt{N}$, as proved by Grover~\cite{Grover:97}. Here, we consider a Hamiltonian implementation through a quantum system 
composed of $n$ qubits, whose Hilbert space has dimension $N=2^n$. We denote the computational basis by the set $\{|i\rangle\}$ ($0\le i\le N-1$). 
Without loss of generality, we can assume an oracular model such that the marked element is the state $| 0 \rangle$. So the implementation of the quantum search 
can be achieved through the projective Hamiltonian
\begin{equation} \label{eq:hamildef}
H(s) = (1-s)(\mathbb{1}-|{\psi_0}\rangle \langle \psi_0|)+s(\mathbb{1}-| 0 \rangle \langle 0 |), 
\end{equation}
where
$|{\psi_0}\rangle = ({1}/{\sqrt N})\sum_{i=0}^{N-1}| i \rangle$,
and $s$ denotes the normalized time $0 \le s \le 1$. By preparing the system in its ground state at time $t=0$ and by considering an adiabatic dynamics, 
it evolves to the corresponding instantaneous ground state at later times. In particular, the system exhibits a first-order 
QPT at $s=1/2$. The ground state energy in terms of the normalized time $s$ reads
\begin{equation}
E(s) = \frac{1 - \sqrt{1-4s(1-s)\overline{N}}}{2},
\end{equation}
with $\overline N = 1-1/N$. For the ground state vector $| {\psi (s)}\rangle$, we obtain 
\begin{equation}
| {\psi(s) }\rangle= \sqrt{a(s)}| 0 \rangle + \sqrt{c(s)}\sum_{i=1}^{N-1} | i \rangle, \label{GS-Grover}
\end{equation}
where we have defined the quantities
$a(s) = \frac{1}{1+(N-1)k_s^2}$, $c(s) = \frac{k_s^2}{1+(N-1)k_s^2}$, and 
$k_s = 1 - \frac{E(s)}{(1-s)\overline{N}}$.    
%%% v2
Note that, in the thermodynamic limit $n \rightarrow \infty$, the structure of the Hamiltonian implies that the LQU can only be non-vanishing at the quantum critical point,  
even though its scaling is nontrivial at finite sizes. This can observed from Eq.~(\ref{eq:hamildef}), 
where both $|0\rangle$ and $|\psi_0 \rangle$ are product states that become orthogonal for $n \rightarrow \infty$. In this
limit, the ground state is $|\psi_0\rangle$ for $0\leq s < 1/2$, with energy $E(s)=s$, while the ground state is $|0\rangle$ for $1/2 < s \leq 1$, with
energy $E(s)=1-s$. At $s=1/2$ the ground state is degenerate. From Eq.~(\ref{GS-Grover}), $| {\psi(1/2) }\rangle$ will be an equal superposition 
of $|0\rangle$ and $|\psi_0\rangle$ for $n\rightarrow \infty$. 
It then follows that $Q=0$ everywhere except at $s=1/2$. 
%%%

In order to determine the scaling at finite size $n$, we consider the density matrix $\rho =  | {\psi(s) }\rangle \langle {\psi(s) }|$ describing the system in the ground state, 
which can be written as 
\begin{equation} \label{eq:rho}
\rho(s) =
\begin{bmatrix}
   a & b & b & \dots  & b \\
    b & c & c & \dots  & c \\
    \vdots & \vdots & \vdots & \ddots & \vdots \\
    b & c & c & \dots  & c
\end{bmatrix} ,
\end{equation}
where $b = \sqrt{a(s)c(s)}$. As we trace out $n^\prime$ qubits of the system, 
the resulting partial density matrix $\rho^\prime(s)$ will be given by
\begin{equation} \label{eq:rhotrace}
\rho^\prime(s) =
\begin{bmatrix}
   a^\prime & b^\prime & b^\prime & \dots  & b^\prime \\
   b^\prime & c^\prime & c^\prime & \dots  & c^\prime \\
    \vdots & \vdots & \vdots & \ddots & \vdots \\
    b^\prime & c^\prime & c^\prime & \dots  & c^\prime
\end{bmatrix},
\end{equation}
where $a^\prime = a +(2^{n'}-1)c$, $b^\prime = b +(2^{n'}-1)c$, and $c^\prime = 2^{n'} c$. We observe that 
$\rho^\prime$ is an $N^\prime \times N^\prime$ matrix, with $N^\prime = 2^{n-n^\prime}$.
Taking the square root from Eq.~(\ref{eq:rhotrace}), we obtain
\begin{equation} \label{eq:mraiz}
\sqrt {\rho^\prime} =
\begin{bmatrix}
   a_r & b_r & b_r & \dots  & b_r \\
    b_r & c_r & c_r & \dots  & c_r \\
    \vdots & \vdots & \vdots & \ddots & \vdots \\
    b_r & c_r & c_r & \dots  & c_r
\end{bmatrix},
\end{equation}
where we have defined
$a_r =\lambda^++\lambda^-$, 
$b_r =\lambda^+\beta^++\lambda^-\beta^-$, and 
$c_r =\lambda^+(\beta^+)^2+\lambda^-(\beta^-)^2$, 
with
$r = \sqrt{4 (b^\prime)^2(N^\prime-1)+(a^\prime - c^\prime(N^\prime-1))^2}$, $\beta^\pm=\frac{c^\prime(N^\prime-1)-a^\prime \pm r}{2 b^\prime(N^\prime-1)}$ and 
$\lambda^\pm=\sqrt{\frac{a^\prime + c^\prime(N^\prime-1) \pm r}{2}}/(1+(N^\prime-1)(\beta^\pm)^2)$.
By rewriting Eq.~(\ref{eq:mraiz}) in block form, we get
\begin{equation} \label{eq:mraizblocos}
\sqrt {\rho^\prime} =
\begin{bmatrix}
   A & B & B & \dots  & B \\
    B^\dagger & C & C & \dots  & C \\
    \vdots & \vdots & \vdots & \ddots & \vdots \\
    B^\dagger & C & C & \dots  & C
\end{bmatrix},
\end{equation}
where $A$, $B$ and $C$ are $2 \times 2$ matrices defined as
\begin{equation}\label{eq:rootblocks}
\begin{array}{c}
 \begin{array}{rl}
A  =
\begin{bmatrix}
    a_r & b_r \\
    b_r & c_r \\

\end{bmatrix},&
B  =
\begin{bmatrix}
    b_r & b_r \\
    c_r & c_r\\

\end{bmatrix} ,

\end{array} \\
\begin{array}{c}
C  =
\begin{bmatrix}
    c_r & c_r \\
    c_r & c_r \\
\end{bmatrix}.
\end{array}
\end{array}
\end{equation}
Now we define $M_i = \sqrt{\rho'} \cdot (\mathbb{1} \otimes \sigma_i)$, where $\sigma_i$ are the $2\times2$ Pauli matrices.
 We now compute the matrix elements $(W_{AB})_{ij} = Tr[M_i M_j]$. This can be analytically performed for arbitrary size $N$, yielding 
\begin{eqnarray}
W_{1,1} &=&2 \left( b_r^2+ c_r a_r +4 q b_r c_r + 2 q^2 c_r^2 \right) ,\nonumber \\
W_{2,2} &=&2 \left(c_ra_r-b_r^2\right),\nonumber \\
W_{3,3} &=&a_r^2 -2b_r^2 +c_r^2 ,\nonumber  \\
W_{3,1} &=& W_{1,3} = 2\left(b_ra_r - b_rc_r + q b_r^2 - q c_r^2\right), \nonumber \\
\end{eqnarray}
with $q=(N^\prime/2-1)$ and the remaining matrix elements of $W$ vanishing. We notice that the matrix $W$ would be left unchanged if we 
interchange blocks $A$ and $B$, taking block $A$ as the first qubit from the left and $B$ as the L remaining qubits.
In order to obtain the LQU, we have to find out the largest eigenvalue of $W_{AB}$ and use it into Eq.~(\ref{LQU}). Since four entries of the matrix $W$ have vanished, this can be done analytically as well.   

\vspace{0.5cm}

\begin{figure}[!ht]
\centering
\includegraphics[scale=0.31]{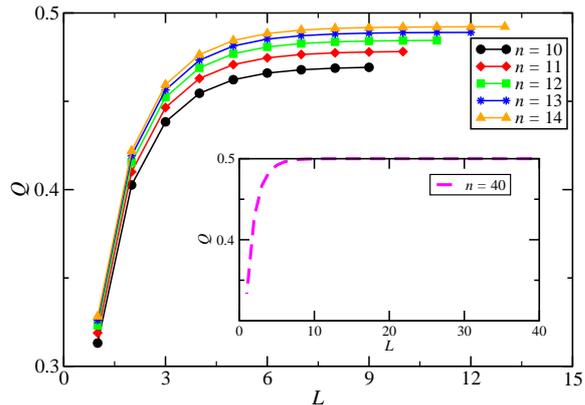}
\caption{(Color online) LQU between a block of length $L$ and a single qubit at the quantum critical point $s=0.5$. 
An exponential saturation as a function of $L$ is observed, with the saturation value enhanced by the system size $n$. The inset shows the curve for $n=40$, which approaches the asymptotic form and can be fit by $Q(L) = 1/2 - a e^{-b L}$, with $a\approx 0.29$ and $b\approx 0.71$.
 }
\label{fig:crit}
\end{figure}
\vspace{0.2cm}

\begin{figure}[!ht]
\centering
\includegraphics[scale=0.31]{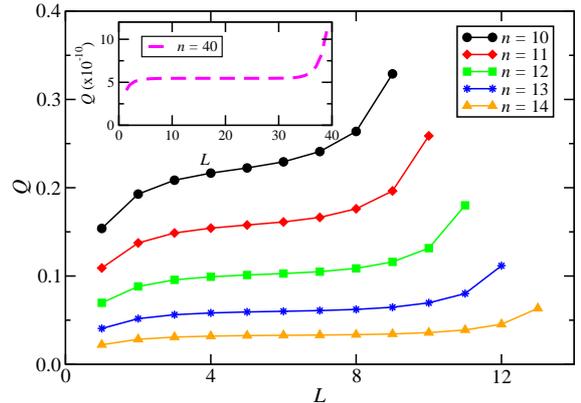}
\caption{(Color online) LQU between a block of length $L$ and a single qubit at the non-critical point $s=0.49$. The LQU tends to vanish as we increase the size $n$ of the system (see Inset).}
\label{fig:norm}
\end{figure}

By taking the LQU between a block of length $L$ and a single qubit, we obtain an exponential saturation of the LQU at the quantum critical point. This is illustrated in Fig.~\ref{fig:crit}, where we plot the LQU as a function of $L$ for $s=1/2$. 
%%% v2
As we can see, the saturation value is enhanced by $n$ and $L$, attaining $Q=1/2$ for $n\rightarrow \infty$ and $L \rightarrow \infty$. 
In this regime, the composite system $AB$ is described by a pure state, with the LQU equivalent to bipartite entanglement as measured by the linear entropy~\cite{Girolami:13}. 
By computing the linear entropy $S(\rho_B)=2 \left(1 - \textrm{Tr} \rho_B^2\right)$ from Eq.~(\ref{GS-Grover}) for $n\rightarrow \infty$ and $L \rightarrow \infty$, 
we directly obtain $S(\rho_B)=1/2$, in agreement with the thermodynamic limit in Fig.~\ref{fig:crit}.
%%%
On the other hand, as displayed in Fig.~\ref{fig:norm}, the behavior of the LQU displays a "tilde" form if the system is driven to a non-critical point. In this case, as the total number of spins $n$ grows, the curve tends to vanish independently of the block length $L$.

\vspace{0.5cm}

\vspace{0.5cm}

\begin{figure}[!ht]
\centering
\includegraphics[scale=0.29]{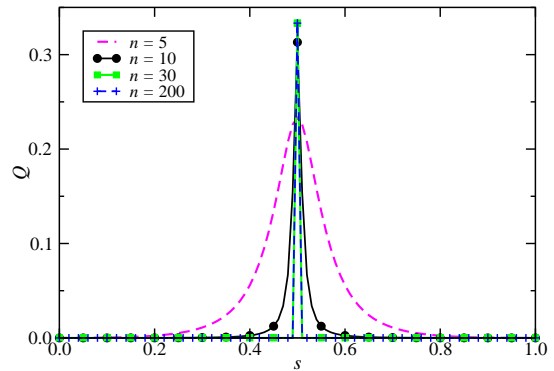}
\caption{(Color online) LQU between a pair of qubits $(L=1)$ as a function of the normalized time $s$ in the ground state. The system shows a sharp behavior at the critical point as a function of the system size $n$. 
In particular, exponential 
convergence of the LQU towards its thermodynamical value is obtained for each value of $s$.}
\label{fig:evo}
\end{figure}

The behavior of the LQU as a function of $s$ for a fixed block size $L$ is also remarkable. Since we have a first order QPT, the LQU itself shows a sharp behavior at $s=1/2$, as shown in Fig.~\ref{fig:evo}. 

\vspace{0.5cm}

\vspace{0.5cm}

\begin{figure}[!ht]
\centering
\includegraphics[scale=0.31]{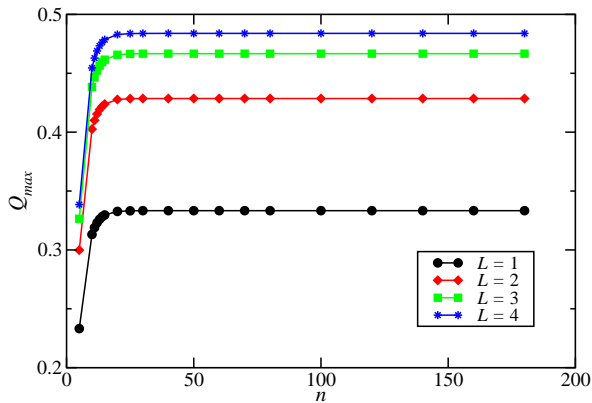}
\caption{(Color online) The maximum of the LQU for distinct sizes $L$ of blocks. }
\label{fig:min}
\end{figure}

Indeed, this is exhibited for a pair of qubits for systems of distinct total number $n$ of qubits. Notice that, for $L=1$, we obtain $Q=1/3$ in the 
 limit $n\rightarrow \infty$. 
%%%v2
This value comes from the structure of $|\psi(1/2)\rangle$ in Eq.~(\ref{GS-Grover}) for $n\rightarrow \infty$, which is an equal superposition of the states $|0\rangle$ 
and $|\psi_0\rangle$.  
%%%
In particular, this maximum value of $Q$, which is obtained at the critical point $s=1/2$, strongly depends on $L$. This is illustrated in Fig.~\ref{fig:min}, where it is shown that the maximum value of the LQU increases along with the total number $n$ of qubits in the system and $L$ for the block, approaching $Q=1/2$ in the limit of both $L$ and $n$ approaching infinity.

\vspace{0.5cm}
\vspace{0.2cm}

%%%%%%%%%%%%%%%%%%%%%%%%%%%%%%%%%%%%%%%%%%%%%%%%%%%%%%%%%%%%%%%%%%%%
\section{The transverse field Ising model} 
%%%%%%%%%%%%%%%%%%%%%%%%%%%%%%%%%%%%%%%%%%%%%%%%%%%%%%%%%%%%%%%%%%%%

\label{sec-ib}

Let us consider now the Ising spin chain in a transverse magnetic field, whose Hamiltonian is given by
\begin{equation}
H_{I}= -J \sum_{i=1}^{n} \left( \sigma^z_i \sigma^z_{i+1} +
\lambda \sigma^x_i \right),
\label{HI}
\end{equation}
with open boundary conditions assumed. Without loss of generality, we will set the energy scale such that $J=1$.
This Hamiltonian is $Z_2$-symmetric and exhibits a second-order QPT from ferromagnetic to paramagnetic state at $\lambda=1$~\cite{Sachdev,Continentino}. 
It can be exactly diagonalized by mapping it to a spinless free fermion model with single orbitals. However, instead of this solution, it is very convenient to consider a treatment based on density matrix renormalization group (DMRG)~\cite{White:92}, since
%%%v2
 $(L+1)$-point correlators and partial density matrices will be the objects of interest for the 
computation of the LQU for block systems.  
In the DMRG scenario adopted here, 
%%%
we truncate the system at each renormalization step 
of the density matrix, keeping a number $M=30$ states for chain sizes up to $48$ sites. This is well justified in the transverse-field Ising model, 
since the neglected states in the DMRG procedure have probabilities of the order of $10^{-10}$.

\vspace{0.5cm}

\begin{figure}[!ht]
\centering
\includegraphics[scale=0.31]{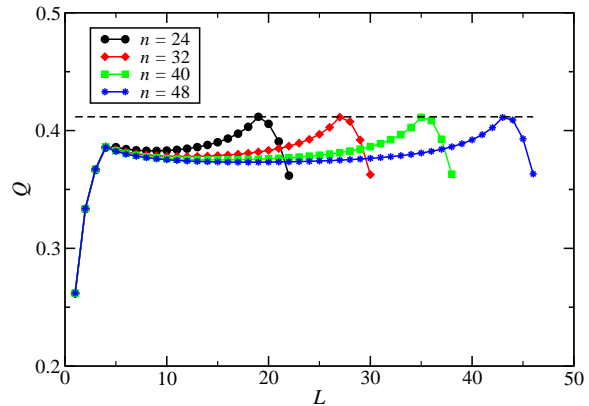}
\caption{(Color online) LQU as a function of the block size $L$ in the transverse-field Ising model at quantum critical point $\lambda=1$. }
\label{fig:ITFcrit}
\end{figure}
\vspace{0.2cm}

\begin{figure}[!ht]
\centering
\includegraphics[scale=0.31]{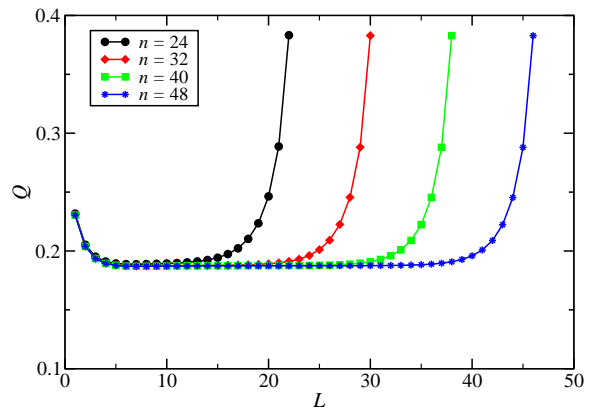}
\caption{(Color online) LQU as a function of the block size $L$ in the transverse-field Ising model at $\lambda=0.8$ (below the critical point). }
\label{fig:ITFbel}
\end{figure}

\begin{figure}[!ht]
\centering
\includegraphics[scale=0.31]{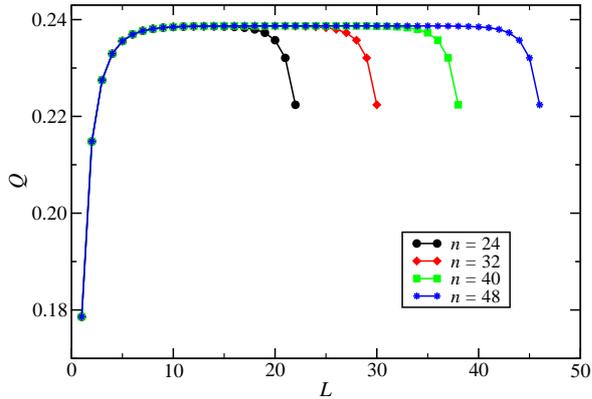}
\caption{(Color online) LQU as a function of the block size $L$ in the transverse-field Ising model at $\lambda=1.2$ (above the critical point). }
\label{fig:ITFabo}
\end{figure}

We consider the behavior of the LQU 
as a function of the block size $L$ for $\lambda=1$, $\lambda=0.8$, and $\lambda=1.2$, which are exhibited in Figs.~\ref{fig:ITFcrit}, 
\ref{fig:ITFbel}, and \ref{fig:ITFabo}, respectively. As we can see, Fig.~\ref{fig:ITFcrit} shows that, at the critical point, the concavity of the 
LQU is undefined. We can also observe that the LQU reaches its maximum value $(Q_{max} \approx 0.41)$ independently of the number $n$ of sites. This is indicated by 
the dashed line in Fig.~\ref{fig:ITFcrit}. On the other hand, concavity is well-defined at off-critical points. Below the critical point, LQU is a convex function of the 
block size $L$, as shown in Fig.~\ref{fig:ITFbel}. Oppositely, LQU is a concave function of $L$ above the critical point, as displayed in 
Fig.~\ref{fig:ITFabo}. 

\vspace{0.5cm}

\vspace{1 cm}

\vspace{0.2cm}

\begin{figure}[!ht]
\centering
\includegraphics[scale=0.31]{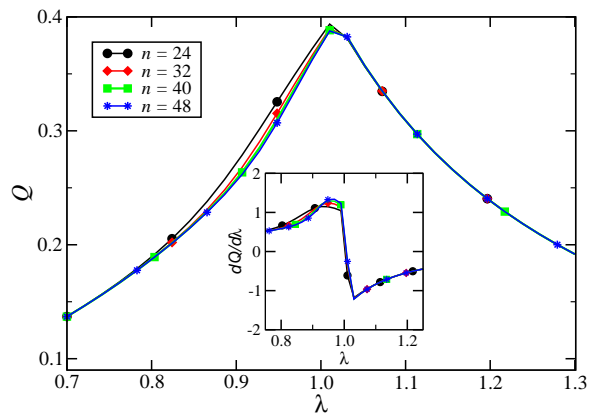}
\caption{(Color online) LQU as a function of the coupling parameter $\lambda$ in the transverse-field Ising model for distinct chain sizes $n$. Inset: 
First derivative of $Q$ with respect to $\lambda$, displaying the characterization of the the QPT.}
\label{fig:ITFlambda}
\end{figure}

The quantum critical point can also be revealed as a function of the coupling parameter $\lambda$ for a 
fixed length $L$. By varying $\lambda$, we can observe a sharp behavior of the LQU around $\lambda=1$ for finite systems. 
This is illustrated in Fig.~\ref{fig:ITFlambda}, where the LQU between a block of $L=10$ sites and a single site is shown for 
distinct lengths $n$ of chains. The inset in Fig.~\ref{fig:ITFlambda} shows the first derivative of the LQU with respect to parameter $\lambda$. As a typical  characterization of the QPT by a quantum discord-like measure~\cite{Sarandy:09}, 
the critical point is identified by the first-derivative of the LQU with respect to $\lambda$. 

\vspace{0.5cm}

\vspace{0.2cm}

\begin{figure}[!ht]
\centering
\includegraphics[scale=0.15]{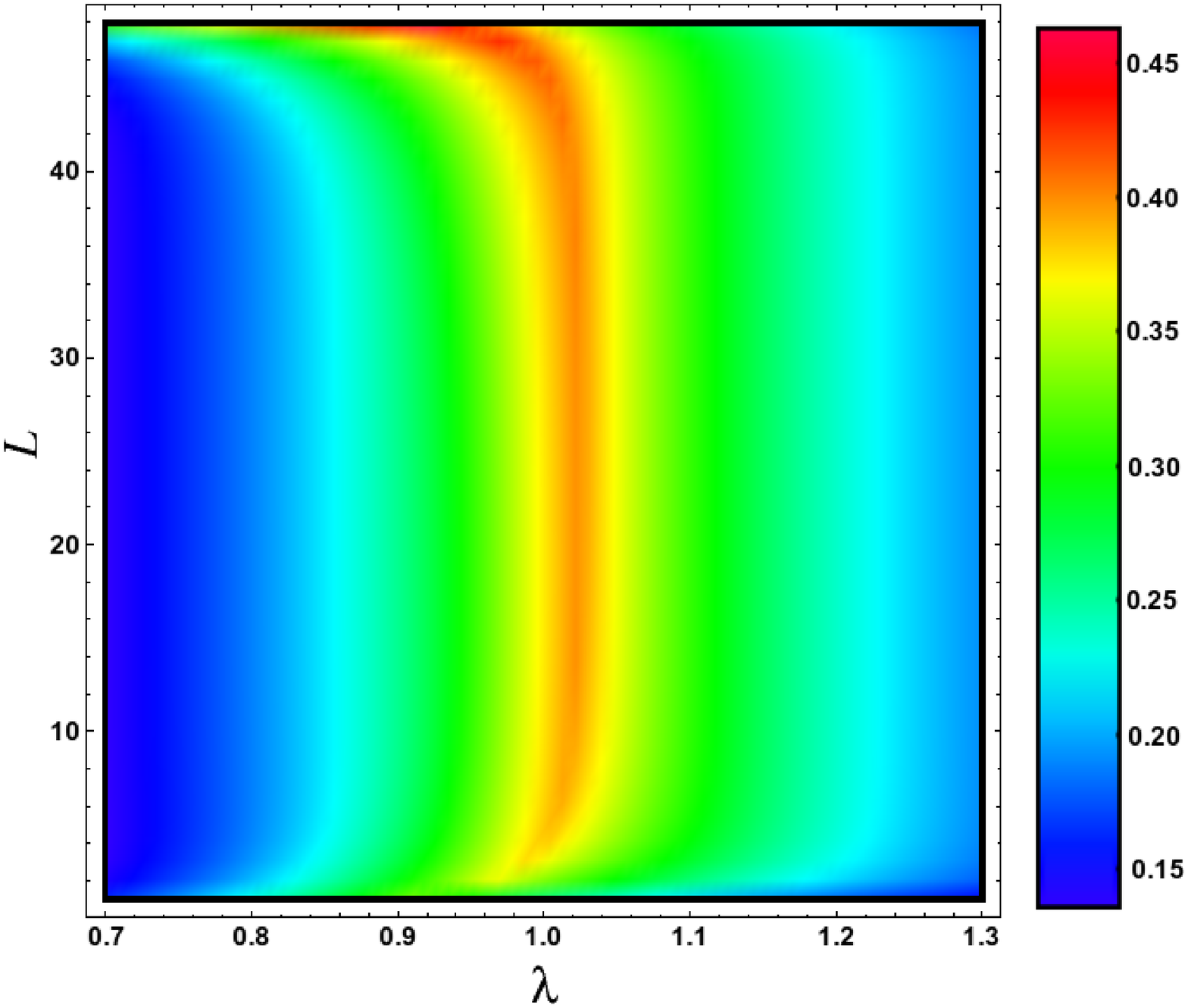}
\includegraphics[scale=0.18]{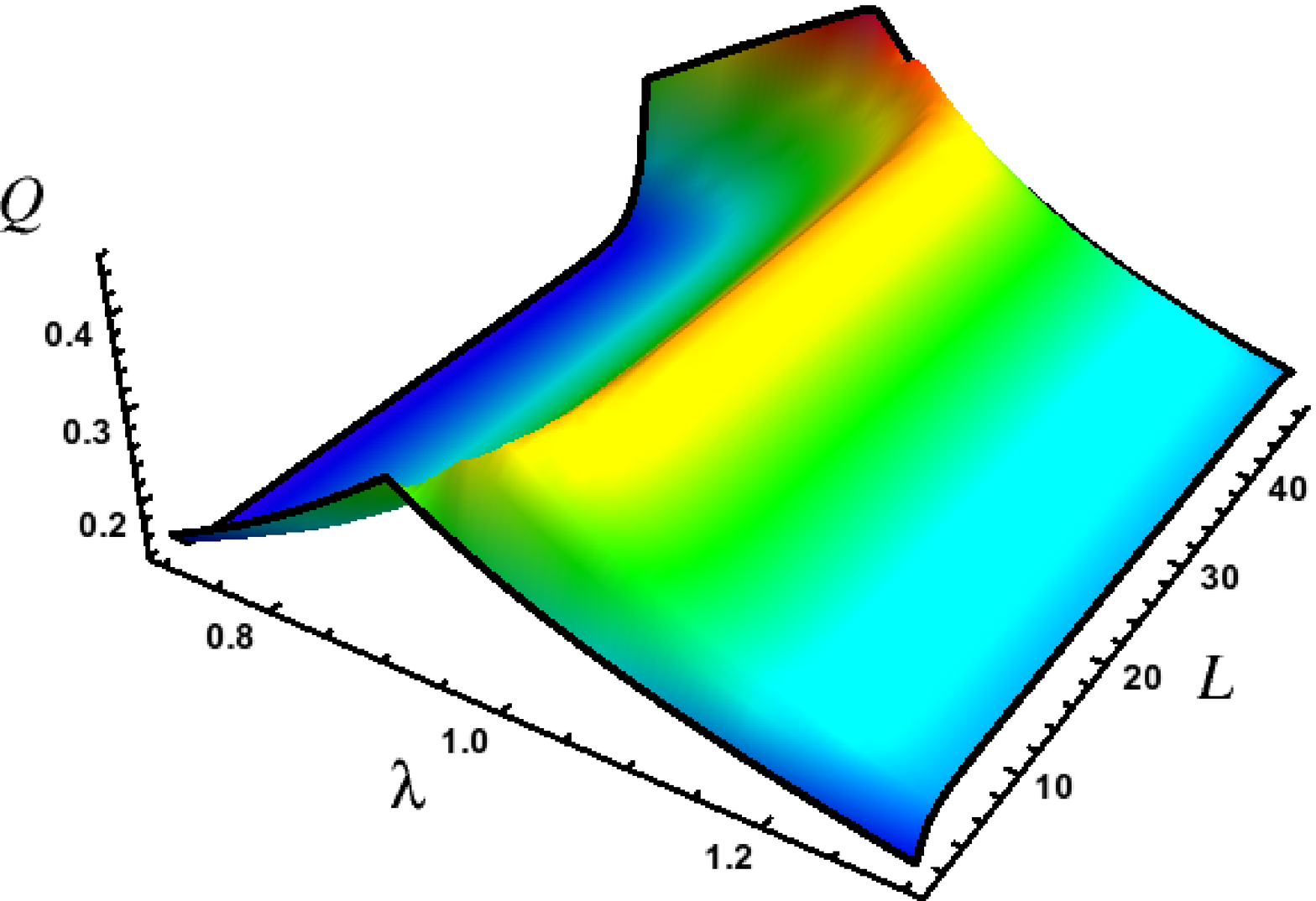}
\includegraphics[scale=0.15]{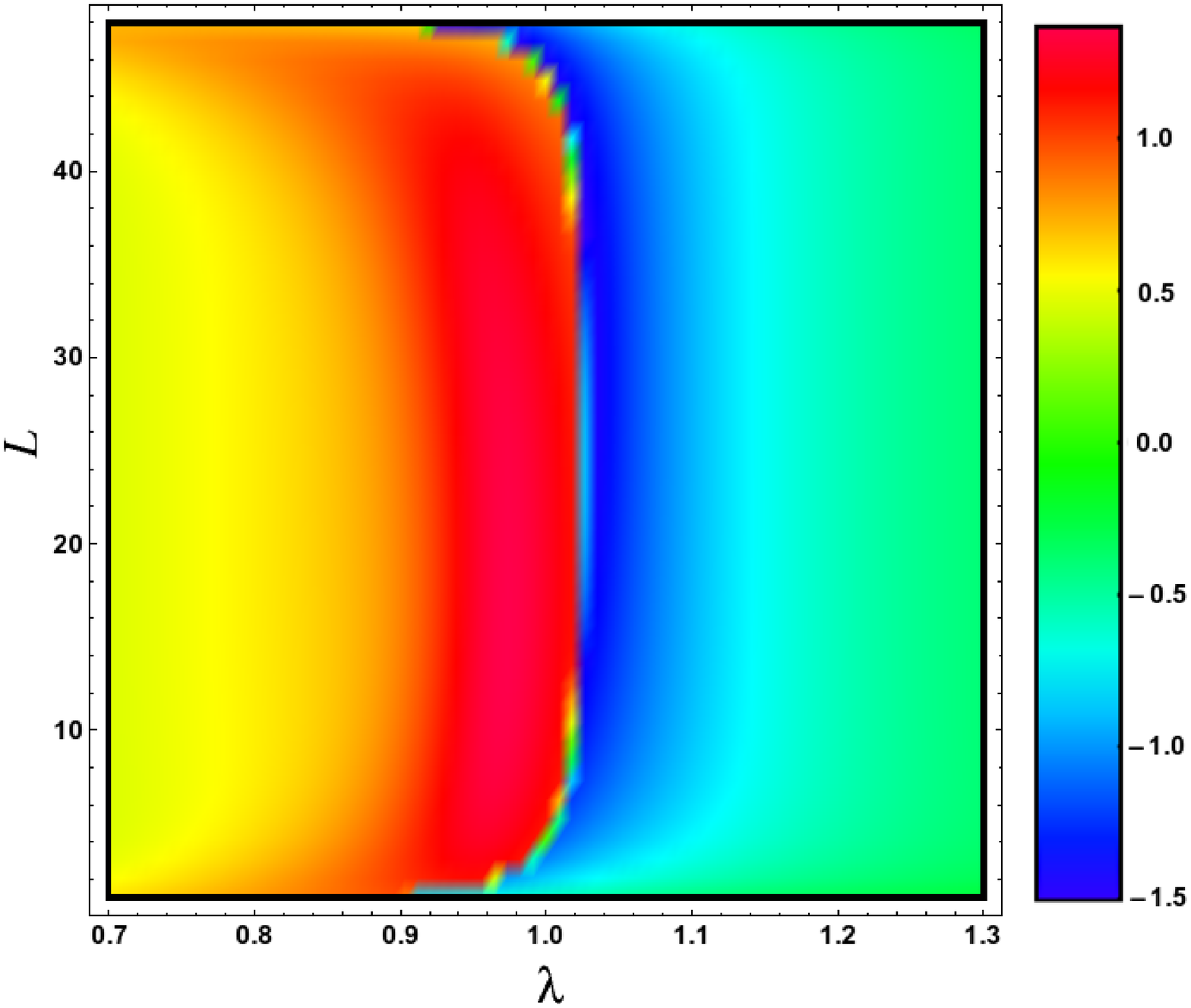}
\includegraphics[scale=0.18]{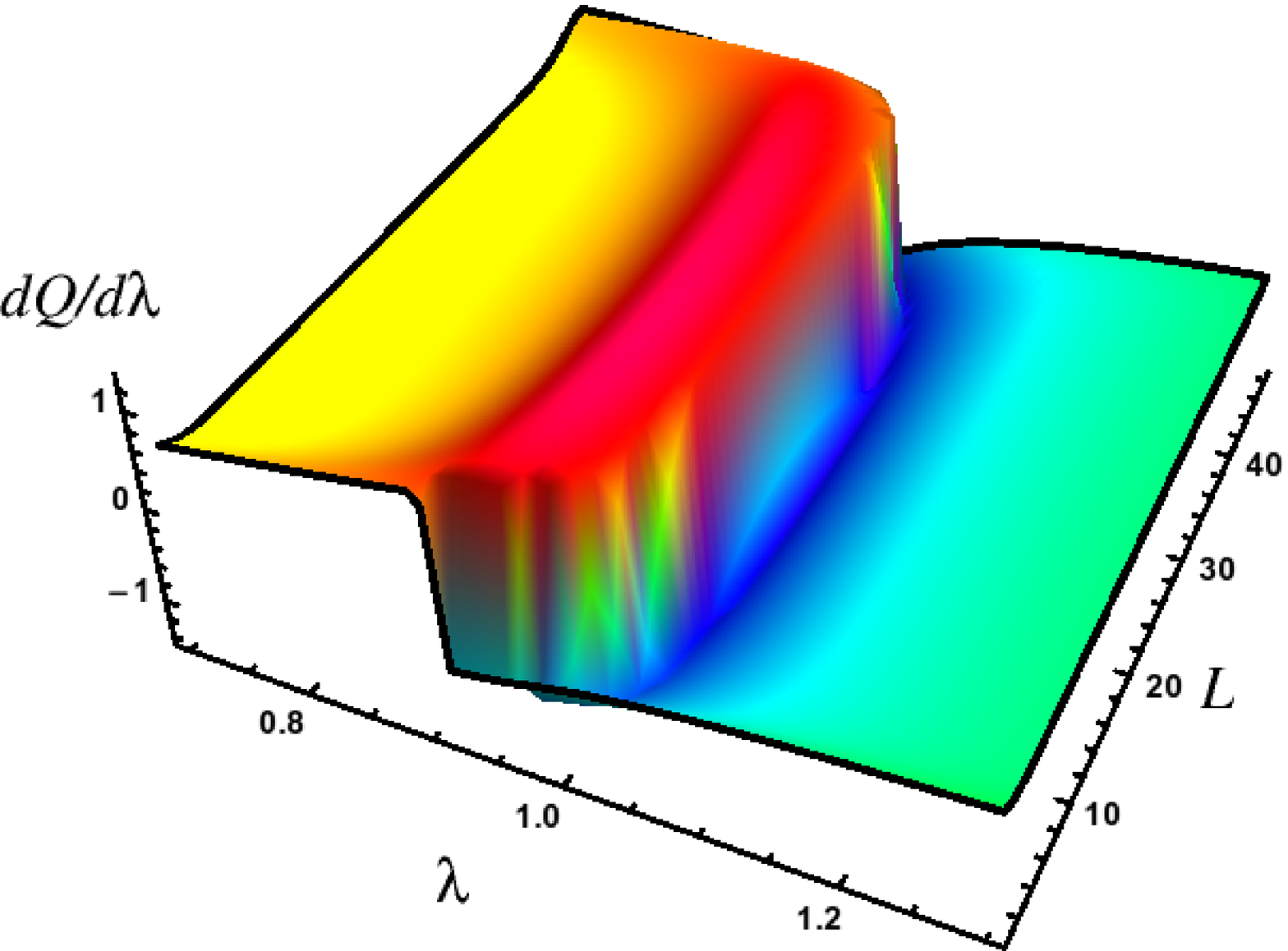}
\caption{(Color online)  LQU and  and its  first derivative as a function of the coupling parameter $\lambda$ and the block size $L$.}
\label{fig:ITFlambdaL}
\end{figure}

In Fig.~\ref{fig:ITFlambdaL}, we show the LQU and its first derivative in the $(\lambda,L)$ plane.
It is observed that around the critical point $(\lambda=1)$, the value of the corresponding LQU on the color scale is approximately $0.4$. The first derivative of LQU is also shown in Fig.~\ref{fig:ITFlambdaL}. The sharp behavior of this quantity characterizes the QPT. The minimum value of the first derivative has a symmetry behavior as a function of block size $L$. 

\vspace{0.5cm}

%%%%%%%%%%%%%%
\section{Conclusion}
%%%%%%%%%%%%%%

In conclusion, we have investigated the scaling properties of the LQU at first-order and second-order QPTs. We have considered the behavior of the LQU in terms of either the block size $L$ or the coupling parameter inducing the QPT. In both cases, the QPTs are precisely identified by the LQU.
%%%v2:
The scaling of the LQU as a function of the size of the block opens the possibility of searching for connections between quantum correlations and finite-size properties of critical systems. 
%%% 
 As a future step, it would be interesting to investigate possible universal properties of LQU, e.g. analyzing its behavior for quantum critical spins chains belonging to distinct universality classes. A relevant point would be a possible scaling related to the central charge of the Virasoro algebra behind the critical chain. 
%%%v2
Other points of interest would be the investigation of the quantum critical behavior for finite temperatures 
as well as the effect of boundary conditions over the scaling of the LQU for systems $D \times 2$. 
%%%
These investigations are left for further research. 

%%%%%%%%%%%%%%
\section*{Acknowledgments}
%%%%%%%%%%%%%%
This work is supported by the Brazilian agencies 
CNPq, CAPES, FAPERJ, and the Brazilian National Institute for Science and Technology of Quantum Information (INCT-IQ).

%\bibliography{bib-file}

\section*{References}

\begin{thebibliography}{10}

\expandafter\ifx\csname url\endcsname\relax
  \def\url#1{\texttt{#1}}\fi
\expandafter\ifx\csname urlprefix\endcsname\relax\def\urlprefix{URL }\fi
\expandafter\ifx\csname href\endcsname\relax
  \def\href#1#2{#2} \def\path#1{#1}\fi

\bibitem{Amico:08}
L. Amico, R. Fazio, A. Osterloh, V. Vedral, 
Rev. Mod. Phys. 80 (2008) 517.

\bibitem{Maruyama:09}
K.~Maruyama, F.~Nori, V.~Vedral,
Rev. Mod. Phys. 81 (2009) 1.

\bibitem{Georgescu:14}
I.~M. Georgescu, S.~Ashhab, F.~Nori,
Rev. Mod. Phys. 86 (2014) 153.

\bibitem{Amico:02}
A.~Osterloh, L. Amico, G. Falci, R. Fazio,
Nature 416 (2002) 608. 


\bibitem{Osborne:02}
T.~J. Osborne, M.~A. Nielsen,
Phys. Rev. A 66 (2002) 032110.

\bibitem{Wu:04}
L.-A. Wu, M.~S. Sarandy, D.~A. Lidar,
Phys. Rev. Lett. 93 (2004) 250404.


\bibitem{Wu:06}
L.-A. Wu, M.~S. Sarandy, D.~A. Lidar, L.~J. Sham,
Phys. Rev. A 74 (2006) 052335.

\bibitem{Vidal:03}
G.~Vidal, J.~I. Latorre, E.~Rico, A.~Kitaev,
Phys. Rev. Lett. 90 (2003) 227902.

\bibitem{Korepin:04}
V.~E. Korepin,
Phys. Rev. Lett. 92 (2004) 096402.

\bibitem{Calabrese:04}
P.~Calabrese, J.~Cardy,
JSTAT 2004 (2004) P06002.

\bibitem{Ollivier:01}
H.~Ollivier, W.~H. Zurek,
Phys. Rev. Lett. 88 (2001) 017901.

\bibitem{Dillenschneider:08}
R.~Dillenschneider,
Phys. Rev. B 78 (2008) 224413.

\bibitem{Sarandy:09}
M.~S. Sarandy,
Phys. Rev. A 80 (2009) 022108.

\bibitem{Maziero:12}
J.~Maziero, L.~Céleri, R.~Serra, M.~Sarandy,
Phys. Lett. A 376(2012) 1540.

\bibitem{Huang:13}
Y. Huang, 
Phys. Rev. B 89 (2013) 054410.

\bibitem{Werlang:10}
T.~Werlang, C.~Trippe, G.~A.~P. Ribeiro, G.~Rigolin,
Phys. Rev. Lett. 105 (2010) 095702.


\bibitem{Girolami:13}
D.~Girolami, T.~Tufarelli, G.~Adesso,
Phys. Rev. Lett. 110 (2013) 240402.


\bibitem{Luo:03}
S.~Luo,
Phys. Rev. Lett. 91 (2003) 180403.

\bibitem{Wigner:63}
E.~P. Wigner, M.~M. Yanase,
PNAS 49 (1963) 910.


\bibitem{Yu:14}
C.~shui Yu, S.~xiong Wu, X.~Wang, X.~X. Yi, H.~shan Song,
EPL (Europhysics Letters) 107 (2014) 10007.

\bibitem{Karpat:14}
G.~Karpat, B.~\ifmmode~\mbox{\c{C}}\else \c{C}\fi{}akmak, F.~F. Fanchini,
Phys. Rev. B 90 (2014) 104431.


\bibitem{Carrijo:15}
T.~M. Carrijo, A.~T. Avelar, L.~C. Céleri,
J. Phys. B: Atomic, Molecular and Optical Physics 48 (2015) 125501.

\bibitem{Li:08}
N.~Li, S.~Luo,
Phys. Rev. A 78 (2008) 024303.

\bibitem{Grover:97}
L.~K. Grover, 
Phys. Rev. Lett. 79 (1997) 325.

\bibitem{Sachdev}
S.~Sachdev, {Quantum Phase Transitions}, 
Cambridge University Press, 2011.

\bibitem{Continentino}
M.~A. Continentino,
{Quantum Scaling in Many-body Systems}, 
World Scientific lecture notes in physics, World Scientific, 2001.

\bibitem{White:92}
S.~R. White, 
Phys. Rev. Lett. 69 (1992) 2863.


\end{thebibliography}

\end{document}